\documentclass{svjour3}                     
\smartqed  
\usepackage{graphicx}
\usepackage{amsmath}
\usepackage{amsfonts}
\usepackage{amssymb}
\usepackage{bm}
\usepackage{subfig}

\newcommand{\beq}{\begin{equation}}
\newcommand{\eeq}{\end{equation}}
\newcommand{\beqa}{\begin{eqnarray}}
\newcommand{\eeqa}{\end{eqnarray}}

\begin{document}

\title{Weinberg's proposal of 1990: A very personal view\thanks{Contribution to Special Issue:
Celebrating 30 years of Steven Weinberg's papers on Nuclear Forces from Chiral
Lagrangians}
}

\titlerunning{Personal view on Weinberg 1990}        

\author{R. Machleidt        
}


\institute{R. Machleidt \at
              Department of Physics, University of Idaho, Moscow, ID 83844, USA\\
              \email{machleid@uidaho.edu}           
}

\date{Received: date / Accepted: date}

\maketitle

\begin{abstract}
My personal encounter with Weinberg's proposal of 1990 was a really  entertaining one:
 My collaborator David Entem and I had embarked to show that Weinberg's idea, though smart and beautiful, was essentially useless in practice (like so many of those genius ideas of the 1980s
 where people claimed to have ``derived the nuclear force from QCD''). 
 However, in trying to do so, we showed the opposite;
  namely, we showed that Weinberg's idea worked better than allowed by any reasonable means.
\end{abstract}

\section*{}
No story of modern physics is more intriguing than the history of the theory of nuclear forces.
It started early, in 1935, just three years after the discovery of the neutron, Yukawa 
made his seminal proposal that a massive particle (later identified with the pion) would mediate the nuclear force~\cite{Yuk35}. This ``pion-theory'' would fail in the 1950's, because
chiral symmetry was not yet an issue. In the early 1960s,
the theory was seemingly rescued by the discovery of heavy (non-strange) mesons,
which were then added to the pion, leading to the one-boson-exchange (OBE) model that was celebrated as a great success~\cite{RMP67}.
In fact, in the 1960s and 70s, many researchers (inluding myself~\cite{HM75,HM76}) believed that meson theory was the fundamental approach to explaining nuclear foces (at least as fundamental as it can get~\cite{Erk74}).
But obviously, single meson exchange leaves out many contributions (like, irreducible multi-meson exchange and contributions involving nucleon resonances~\cite{HM77}) that cannot be
argued away and, so, in the 1970s and 80s, researchers tried to go beyond the OBE model.
 Most notably, this was pursued by 
  the Paris group applying dispersion relations~\cite{Vin79,Lac80} and
our team at the University of Bonn using field theory.
The Bonn model, published in 1987 in its final form~\cite{MHE87}, turned out
to be amazingly quantitative and, thus, seemed to confirm that meson theory was, indeed, {\it the} adequate approach to nuclear forces.

But the 1980s also became a period of change.
Nuclear physicist started to notice that they could not ignore anymore
what particle physicists had promoted already for a decade; namely, that quantum 
chromodynamics (QCD) was the fundamental theory of strong interactions
(and not some sort of meson theory).
Thus, a new fashion emerged, namely, to ``derive the nuclear force from QCD.''
Noticing that the problem is still unsolved 40 years later,
it is not surprising, in hindsight, that only rough models were generated during the 1980s.
The approaches were typically based on the quark models of the nucleon that had started to float around,
like relativistic bag models (MIT bag~\cite{Deg75}), non-relativistic potential models~\cite{IK77},
chiral bags (Little Bag~\cite{BR79}, Cloudy Bag~\cite{Tho83}), 
from which QCD-inspired nucleon-nucleon 
 interactions were 
 derived~\cite{Tar77,OY80,Har81,Fae82,MI83,FF83,OY84,Mor84,Won86,MW88,TSY89,Fer93}.
Besides quark models, also an interesting alternative was offered, namely, the
Skyrme model~\cite{Vin85,JJP85,ZB86,MZ86,WW92}.

In spite of the large number and variety of models that were developed in the course of the 1980s~\cite{Won86,MW88},
it is fairly easy to summarize their general features.
Essentially all models generated some sort of short-range repulsion
 to which the pion was then added as a tail (mostly `by hand,' motivated by the chiral bags).
Unfortunately, the nuclear force is a little bit more sophisticated than just 
one-pion-exchange (OPE) plus a hard core. The intermediate-range attraction was difficult to create in all models, not to mention the spin-orbit force.
Therefore, in some models, the sigma boson---the most despised ingredient of the outdated 
OBE model---suddenly became the lifesaver of the ``QCD approach.''
Watching these developments throughout the 1980s, I had grown critical of 
 all those ``derivations of the nuclear force from QCD.''
Of course, as a matter of honor, all QCD-inspired models claimed some success;
 but most of them
 consisted only of QCD-inspired words with little to show in 
 quantitative terms---unless some mesons had accidentally sneaked in through the backdoor.

Then came the year of 1990 and the paper by Weinberg~\cite{Wei90} (that we are here to celebrate),
in which Weinberg proposed to use chiral effective Lagrangians to study the forces among 
nucleons in terms of powers of the nucleon momenta (chiral perturbation theory).
I found the paper well written and easy to understand (including Weinberg's other papers 
on this topic~\cite{Wei91,Wei92,Wei79}).
However, based upon my observations during the 1980's, I had the prejudice that 
this may be just another
one of those ideas that sound so promising, but were of no practical value---that is,
unable to lead
to an accurate nucleon-nucleon
($NN$) potential to be used in reliable miscroscopic nuclear structure calculations.

 The Weinberg paper of 1990 set off a sequence of actions. 
 It started in the early 1990s with the work by van Kolck and coworkers~\cite{ORK94,ORK96}, who constructed the first chiral $NN$ potenial at next-to-next-to-leading order (NNLO) in configuration space.
 For the inexperienced person the phase shift fits obtained in that work may have 
 given a ``satisfactory'' impression.
  But based upon my comprehensive experience with $NN$ potentials and phase shifts,
 it was immediately clear to me that the fits were poor---meaning that
 the $\chi^2$ in regard to the $NN$ data would have come out very large. Therefore,
  the potential was unsuitable for applications in miscrocopic nuclear structure.
   (Nevertheless, the initial work by van Kolck is of principal historical significance,
   since it marked the starting point.)
 
 The next major step towards developing chiral nuclear interactions was conducted in 1997 
 by Kaiser {\it et al.}~\cite{KBW97}, who derived the 
 two-pion exchange (2PE) contributions up to NNLO in momentum space and in dimensional regularization. This provided the needed mathematical expressions for future potential constructions in momentum space.
  They also calculated the phase shifts in
 peripheral partial waves where their OPE plus 2PE contributions showed
  the right trends, but nothing was quantitative.
 
In the step that followed, Epelbaum {\it et al.} in 1998/2000 picked up the Kaiser expressions for the chiral pion-exchanges up to NNLO
and complemented them with the NNLO contact terms, thus, creating the first momentum space potential that also included the lower partial waves~\cite{EGM98,EGM00}. Again, the 
``well described'' phase shifts
weren't that well described, and the $NN$ potential was insufficient
for quantitative applications.
 
 By now, 10 years had elapsed since Weinberg's proposal, and the chiral potential movement started to
 pick up momentum in spite of its poor predictive power, of which most people were not aware.
  This alarmed me because I became concerned that the nuclear physics community
 might be misled on a crucial issue---similarly to what had happened throughout the 1980s.
And so, in 2001, I decided to take a close look at the chiral approach
to nuclear forces to (potentially) show that,
 in spite of great physics arguments,
 this approach cannot produce a potential of sufficient quality for use in nuclear structure.
 
 I started out by assembling the 2PE expressions by Kaiser up to NNLO with the second-order
 contacts to construct a NNLO chiral $NN$ potential.
 And, yes, what came out was pretty poor, certainly far too poor to 
 be of any practical use. 
 I saw what I had expected and predicted. I found confirmed that
 also this was just another one of those 1980s stories.
 
 But once started and having all the tools at my disposal, I became  curious and took a closer look at various aspects of the problem.
 
 The first thing I noticed was that contacts at NNLO (second order) contribute only in $S$ and $P$ waves. 
 Note that, in this theory, the contacts are the short-range part of the force.
 Thus, at NNLO, the theory assumes that no short-range corrections are needed in $D$ (and higher) waves.
 But from the meson theory of nuclear forces, on which I had worked for many years,
 I knew that this was wrong.
 In meson theory, the short-range contributions come from $\omega$ and $\rho$ exchange,
  which contribute very noticeably in $D$ waves. Moreover, the 2PE contribution
 typically leads to too much attractions in $D$ waves no matter what theory is applied
 to derive the 2PE contribution: dispersion theory (Paris potential~\cite{Vin79}), meson field theory (Bonn potential~\cite{MHE87}), or chiral perturbation theory~\cite{KBW97,EM02a}.
 OPE plus 2PE describe {\it peripheral} $NN$ scattering and they do it well
 in any theory~\cite{EM02}. But $D$ waves are not peripheral. 2PE plus moderate short-range
 repulsion is needed to get those $D$ waves right in any theory.
 
 The question then is:
 How do we get repulsion into those $D$ waves? Well,  working with a perturbation theory, 
 that is easy: just go up in the order. The contact operators of fourth order contribute in $D$ waves.
 So my collaborator David Entem and I derived the contacts up to fourth order (N$^3$LO) and included them.
 Note also that, at second order, there are only a total of nine contacts, while at fourth order there are 
  15 additional terms, leading to a total of 24 contacts (implying 24 free parameters).
  The Nijmegen phase shift analysis~\cite{Sto93} needed 35
 parameters to fit the $np$ data and high-accuracy potentials~\cite{Sto94,WSS95,Mac01} typically use in the
 order 40 parameters. When conventional theories need 35-40 parameters, 
 then using 24 parameters is much more realistic than having just nine.
And so we included the contacts up to fourth order and obtained an excellent fit~\cite{EM02a}.
 
As a last measure, Entem and I also added in the 2PE contributions up to 
N$^3$LO (fourth order)~\cite{EM02,Kai01,Kai02},
to be consistent with the order up to which contacts are needed.
 But the  N$^3$LO 2PE wasn't crucial for the good outcome of the final potential. 
 In fact, when we added all
 the many and complicated N$^3$LO 2PE contributions, one-by-one, we prayed that, in the end, 
 they wouldn't change substantially the 2PE at NNLO.
  Our prayers were heard.
 And so it happened that Entem and I constructed a consistent chiral N$^3$LO 
 potential~\cite{EM03} that was
  as quantitative as any of those high-accuracy potentials of the 1990s and, therefore, 
  most suitable for {\it ab initio}
 nuclear structure calculations---as proven in many applications of this potential in the following years.
 
 In summary, my personal adventure with Weinberg's proposal of 1990 was quite
 engaging. With all due respect to Weinberg's reasoning, 
   I suspected that Weinberg's idea would be  of little practical value.
   This skepticism was based upon the experiences with the many failed attempts of the 1980s
   to ``derive the nuclear force from QCD,'' and
   I felt that there was no need to repeat history.
 However, in trying to prove my point, my collaborator Entem and I showed the opposite;
  namely, we showed that Weinberg's idea worked beyond
  any reasonable expectation~\cite{EM03}. \\
 No kidding: when this happens, research is fun.

\begin{acknowledgements}
This work was supported in part by the U.S. Department of Energy
under Grant No.~DE-FG02-03ER41270.
\end{acknowledgements}

\end{document}